\documentclass[12pt]{article}
\begin{document}

\begin{center}
{\bf Summary of the Working Group on Spin Physics\footnote{Summary 
talk of the Working Group on Spin Physics at DIS2003, 
XI International Workshop on Deep Inelastic Scattering,
St. Petersburg, 23-27 April 2003}}\\
\vskip 0.8cm
{\sf Mauro Anselmino$^2$, Gerard van der Steenhoven$^3$}
\vskip 0.5cm
{\it $^2$Dipartimento di Fisica Teorica, Universit\`a di Torino and \\
          INFN, Sezione di Torino, Via P. Giuria 1, I-10125 Torino, Italy}\\
\vspace{0.3cm}
{\it $^3$Nationaal Instituut voor Kernfysica en Hoge-Energiefysica (NIKHEF),\\
P.O. Box 41882, 1009 DB Amsterdam, The Netherlands}
\end{center}

\vspace{1.5cm}

\begin{abstract}
\noindent 
A summary is given of the experimental and theoretical results presented 
in the working group on spin physics. New data on inclusive and 
semi-inclusive deep-inelastic scattering, combined with theoretical studies 
of the polarized distribution functions of nucleons, were presented. 
Many talks addressed the relatively 
new subjects of transversity distributions and generalized parton 
distributions. These distributions can be studied by measuring
single spin asymmetries, while partonic intrinsic motion and models
of new spin dependent distribution and fragmentation functions are needed to
obtain the corresponding theoretical description. These
subjects are not only studied in deep-inelastic lepton scattering, but also
in polarized proton-proton collisions at RHIC. A selection of results that
have been obtained in these experiments together with several associated 
theoretical ideas are presented in this paper. In conclusion, a brief sketch 
is given of the prospects for experimental and theoretical studies of the spin 
structure of the nucleon in the coming years.
\end{abstract}

\section{Introduction} 

Experimental studies of the spin structure of the proton and the theoretical 
interpretation of the obtained results, are mostly aimed at identifying the 
various carriers of angular momentum inside the proton, and measuring their 
individual 
contributions to its total spin. In the past most studies were based on 
inclusive spin-dependent deep-inelastic lepton scattering experiments, which 
showed, in the QCD parton model interpretation, that the spin of the quarks 
can only account for about 30\% of the spin of the nucleon~\cite{EMC88}.
In recent years many more probes of the spin structure of the nucleon have 
become available: semi-inclusive deep-inelastic scattering to study the 
polarization of individual quark flavors, photon-gluon fusion to measure the
gluon polarization, exclusive processes giving (indirectly) access to
possible orbital angular momentum contributions, and single spin asymmetries
to investigate the transverse spin distribution. This rich variety of
possible probes of the nucleon spin is reflected by the many experiments 
underway or in preparation that (will) provide experimental data on
this subject. At this workshop data from HERMES at DESY, COMPASS at CERN,
Hall A and B at JLab, and STAR and PHENIX at RHIC were presented, thus
illustrating the expansion of the field. 

A parallel effort has been made on the theoretical side. Starting 
a long time ago from the unexpected data of Ref.~\cite{EMC88} (the so-called
`proton spin crisis'), our understanding of the spin structure of the nucleon
has by now greatly improved, stressing the role of 
quark and gluon polarization, intrinsic partonic motion and orbital angular 
momentum. New generalized QCD analyses of structure functions and spin effects
involving parton transverse motion, and (among others) quark-gluon correlations
have been proposed, which open the way to a new interpretation of existing 
data and suggestions for new measurements of spin obervables.   
More in particular, the increased experimental efforts to study the spin 
structure of the nucleon partly follow two such theoretical developments:
\begin{itemize}
\item{The (re-)introduction of generalized parton distributions (GPDs) 
for the description of partonic correlations provides a unified framework for 
the description of a wide range of different experiments. 
By taking the first moment of a combination of certain GPDs, for example, 
information 
can be obtained on the total angular momentum $J_{q}$ of the quarks in the 
nucleon~\cite{Ji}. By comparing measured values of $J_{q}$ to the total spin 
carried by quarks $\Delta\Sigma_q$, information might be obtained on the 
orbital motion of quarks in the nucleon.}
\item{It is predicted by both the chiral-soliton (instanton) 
model~\cite{Chiral} and lattice gauge calculations~\cite{Lattice} that the 
tensor charge of the nucleon ($\delta\Sigma_q$), which can be derived from 
the {\it transversity} distribution function $h_1(x)$ of the nucleon, is 
considerably larger than the longitudinal quark spin contribution 
$\Delta\Sigma_q$, which is derived from data on the longitudinal spin-dependent
structure function $g_1(x)$. The difference between $\delta\Sigma_q$ and 
$\Delta\Sigma_q$ is caused by the absence of gluon-splitting contributions 
in the transverse case, which is also expected to result in a 
relatively weak $Q^2$ dependence of $h_1(x)$. Unfortunately, inclusive 
deep-inelastic scattering cannot be used to measure $h_1(x)$ as it is a 
chirally odd quantity. In semi-inclusive DIS information on $h_1(x)$ can 
be obtained if a process is identified that is governed by
a chirally odd fragmentation function~\cite{MT}.}
\end{itemize}
At the workshop many talks were devoted to first results and/or new plans 
to measure either the generalized parton distributions or the transversity
distribution. 

Given all these developments the presentations submitted to the working
group on spin physics were split in the following categories:
\begin{enumerate}
\item{Inclusive spin structure \& QCD analyses}
\item{Semi-inclusive DIS \& spin-flavor decompositions}
\item{Gluon polarization \& charm production}
\item{Exclusive reactions \& generalized parton distributions}
\item{Transversity \& (single) spin asymmetries}
\end{enumerate}
The same five categories are used in the next section, where the experimental 
and theoretical results presented in the working group are summarized.
In the final section of the paper some perspectives for spin physics are
sketched.

\section{New experimental and theoretical results}

In the following five subsections the most important results presented at 
the workshop are summarized. It is noted that unavoidable
personal biases are introduced whenever a rich collection of new results
has to be summarised. For a full collection of subjects and results
discussed at the workshop the reader is referred to the individual 
contributions appearing elsewhere in these proceedings. To avoid unnecessary 
duplication, the figures representing the various new results are not
reproduced in the subsections below.

\subsection{Inclusive spin structure \& QCD analyses}

Following last years release of very precise $g_1^d(x)$ data by the HERMES
Collaboration, this year the collaboration presented QCD fits of all available 
$g_1(x)$ data~\cite{DeNardo}. Using two different techniques, fairly precise
(and consistent) singlet and non-singlet parton distributions were obtained.
It has also been possible to extract values for the polarized gluon 
distribution $x \Delta G(x)$ with relatively large margins of uncertainty. 
The resulting first moment $\Delta G$ of the polarized gluon distribution is 
still largely uncertain, ranging from about +0.1 to +1.0, but most likely 
positive.

Exploiting the high luminosity available at JLab, the Hall-A collaboration
presented new high-$x$ data obtained in inclusive polarized electron scattering
from a polarized $^3$He target~\cite{McCormick}. The resulting values for 
$A_1^n(x)$ cover the $x$-domain from 0.3 to 0.6 and have superior precision as 
compared to existing data. If such data are extended to even higher $x$-values 
in the future it will be possible to determine the limiting value of 
$A_1^n(x)$ at $x \rightarrow$ 1 for which widely different predictions exist.

In the deuteron the quarks can be polarized even if the deuteron itself is
unpolarized. Such a non-zero quark polarization can be caused by nuclear 
interaction effects, and needs to be accounted for when the spin-dependent
structure function of the neutron $g_1^n(x)$ is extracted from polarized
deep-inelastic scattering data on the deuteron. The effect of a possibly
small quark polarization in an unpolarized deuteron target can be studied by 
measuring the tensor structure function $b_1^d(x)$, which can be determined 
from deep-inelastic scattering
experiments if a tensor polarized deuterium target is used. Such measurements 
were carried out for the first time by the HERMES Collaboration at 
DESY~\cite{Stancari}. The data presented show that $b_1^d(x)$ is small, being 
consistent with zero for $x >$ 0.1 and slightly positive at lower values 
of $x$.

>From the theoretical point of view the agreement between the QCD predictions 
on the evolution of the polarized structure function $g_1^p(x, Q^2)$ and the
data is very good; however, the amount of available data and the explored 
kinematical regions, if compared with the analogous information on the 
unpolarized distribution functions, is still rather poor. The Bjorken sum 
rule on the difference between the first moment of $g_1^p(x, Q^2)$ and  
$g_1^n(x, Q^2)$ is well obeyed, while the few data on $g_2(x, Q^2)$
do not yet allow a definite conclusion about the validity of the 
Burkhardt-Cottingham sum rule, although indicating its possible violation 
\cite{e155}. 

Some theoretical issues are still open concerning heavy quark and higher twist 
contributions, the small-$x$ behavior, radiative corrections and modeling of 
polarized distribution functions. New results were presented on the twist-2 
heavy quark contribution to $g_2(x, Q^2)$ within a covariant parton model, 
obeying the Wandura-Wilczek and Burkhardt-Cottingham relations \cite{blumlein}.
A model independent determination of higher-twist contributions to 
$g_1(x, Q^2)$ was presented by Alexander Sidorov \cite{sidorov}, and another 
study of higher-twist effects, within an infrared-renormalon model, by 
Andrei Kataev \cite{kataev}. The small-$x$ behaviour of the singlet 
$g_1$ function, taking into account effects due to the running of the strong 
coupling constant $\alpha_s$, was obtained by Boris Ermolaev \cite{ermolaev}.
Radiative corrections were discussed by Eduard Kuraev \cite{kuraev} and 
statistical modeling of polarized distributions by Jacques Soffer 
\cite{soffer}. Finally, the importance of the Bloom-Gilman duality
when discussing inclusive polarized parton distributions 
was discussed by Alessandra Fantoni \cite{fantoni}.
      
\subsection{Semi-inclusive DIS \& spin-flavor decompositions}

In semi-inclusive deep-inelastic scattering from polarized targets a (possibly
identified) hadron is observed in coincidence with the scattered lepton.
The hadron serves to tag the flavour of the quark struck in the scattering 
process. Assuming that the fragmentation process is sufficiently well known 
the observed double spin asymmetries for several identified hadron species 
can be converted to helicity distributions for individual quark flavours.
The results of such an analysis were presented by 
Marc Beckmann~\cite{Beckmann} 
with supporting information provided by Achim Hillenbrand~\cite{Hillenbrand}
on the treatment of the fragmentation process. The data, which were derived
from asymmetries measured at HERMES, demonstrated that each of the helicity
densities for $\bar{u}$, $\bar{d}$ and $s$ sea quarks are consistent with zero.
Also the helicity difference distribution $\Delta\bar{u} - \Delta\bar{d}$
was shown to be consistent with zero, which is remarkable as existing
calculations~\cite{Soliton,GRSV} suggest a positive value for this
distribution.

In the nearby future more data on flavor-separated helicity distributions can 
be expected from Hall-A (at high $x$ and using some -- sofar relatively 
strong -- simplifying assumptions~\cite{McCormick}) and COMPASS (at low 
$x$~\cite{Marchand}). It is of interest to note that the polarized $p-p$ 
collisions at RHIC will provide a completely independent measure of the 
flavour-separated quark helicity distributions using the parity violating 
$W^+$ and $W^-$ production channels. Simulated data for the PHENIX experiment 
at RHIC indicate that a high statistical precision can be obtained for the 
spin-dependent distributions $\Delta u$, $\Delta d$, $\Delta \bar{u}$ and 
$\Delta \bar{d}$~\cite{Makdisi}. 

\subsection{Gluon polarization \& charm production} 

Gluons are expected to play an important role in explaining the spin structure
of the nucleon. At present, only a single measurement exists on the gluon
polarization in the proton~\cite{Hermes-g}, but this measurement suffers from
large statistical and systematic uncertainties. Hence, new dedicated 
experiments are needed to measure the gluon polarization $\Delta G$ much 
more precisely. At the workshop three such experiments were discussed:

\begin{itemize}
\item{The COMPASS Collaboration~\cite{Marchand} intends to measure 
$\Delta G$ by identifying photon-gluon fusion processes in deep-inelastic 
muon scattering through charm production or the observation of high 
$p_T$-pairs. The COMPASS experiment was successfully commissioned in 2002, 
and first longitudinally polarized data have been collected, but no results 
on $\Delta G$ are yet available.}
\item{The RHIC-Spin Collaboration~\cite{Makdisi} will measure double spin
asymmetries in polarized $p-p$ collisions. Very precise values of $\Delta G$ 
can be obtained if it is possible to identify prompt photons in the 
photon-gluon Compton process. No data have been collected yet, but it is 
very encouraging that it has been demonstrated to be possible to inject, 
ramp and maintain a polarized proton beam in RHIC.}
\item{The SLAC E161 Collaboration~\cite{Griffioen} has prepared a proposal
to measure $\Delta G$ by identifying photon-gluon fusion events through
open charm production. At present it is unclear when the experiment will run.}
\end{itemize}

\subsection{Exclusive reactions \& generalized parton distributions}

While a precise experimental knowledge of the Generalized Parton Distributions 
(GPDs) would give essentialy complete information on the structure of the 
nucleons, the usual inclusive structure functions and elastic form factors 
only provide limited information as they represent limiting cases of the GPDs. 
This argument explains the importance of the GPD framework, as was discussed 
in the introductory talk on this subject by Maxim Polyakov \cite{polyakov}.  

Access to the Generalized Parton Distributions (GPDs) is obtained by 
identifying exclusive reactions in high-energy lepton-induced experiments. 
By making use of a variety of polarization states, targets and final-state 
hadrons the experiments are sensitive to different combinations of 
GPDs~\cite{Belitsky}. Hence, many different observables need to be extracted 
from the experimental data in order to be able to map the GPDs. The HERMES 
Collaboration~\cite{Amarian} made a start with this effort by determining 
various asymmetries from their data. In the channel corresponding to exclusive 
photon production (or Deeply Virtual Compton Scattering) beam-spin asymmetries,
beam-charge asymmetries, and target-spin asymmetries were shown. In each case 
non-zero asymmetries were reported, but the statistical precision in the 
various kinematical distributions is still rather modest. Single target-spin 
asymmetries were also shown for exclusive 
$\rho^0$ and $\pi^+$ production data~\cite{Elsenbroich}. Also in this case 
the observed asymmetries are small and the statistical precision is still 
limited. On the other hand, the successful determination of each of these
asymmetries on the basis of the relatively low-luminosity HERMES data, implies
that it is in principle possible to measure the various GPDs if in due time
dedicated higher-luminosity experiments become available.

\subsection{Transversity \& (single) spin asymmetries} 

The transversity distribution is the only leading-twist quark distribution for
which no experimental data are available. It is a chirally odd quantity 
and cannot be accessed in fully inclusive DIS, where it cannot be combined with
a suitable chiral-odd partner to yield an observable cross section.
On the other hand, it can couple, in semi-inclusive DIS or in high-energy
proton-proton collisions, to other chiral-odd functions, giving rise to 
interesting and unexpected single spin effects \cite{teryaev}.    
 
One way of obtaining information on the transversity distribution $h_1(x)$ is
by measuring single target-spin asymmetries in semi-inclusive deep inelastic
scattering on transversely polarized targets. Such measurements have been
started both at COMPASS~\cite{Pagano} and HERMES~\cite{Schnell}, but analysis
results are not yet available. In the absence of transverse polarization data, 
existing longitudinally polarized data have been used to determine the 
single-target spin asymmetry $A_{UL}$; because of the -- on average -- 
non-zero lepton scattering angle the measured asymmetry can originate from a 
small ($\approx$ 15\%) transverse spin component. At the workshop the HERMES 
data on 
$A_{UL}$ for both the proton and the deuteron were presented~\cite{Schill}. 
The measured asymmetries are well reproduced by two independent calculations 
that include rough estimates for $h_1(x)$ and the chiral-odd (Collins) 
fragmentation function $H_1^{\perp}$~\cite{Ma,Efremov}. These
calculations only include the Collins effect, {\it i.e.} the process 
sensitive to $h_1(x)$, but ignore possible contributions from the Sivers 
effect, {\it i.e.} the process involving the $k_{\perp}$-dependent structure 
function $f_{1T}^{\perp}$ in combination with the usual chiral-even 
fragmentation function. Transversely polarized data are needed to separate 
these two effects. Their relative importance has been discussed by Anatoli
Efremov \cite{efremov2}. A general discussion on azimuthal asymmetries in 
semi-inclusive DIS, including model estimates of the Collins function,  
was presented by Karo Oganessyan \cite{ogan}.

A combination of the Collins and Sivers effect is also expected to be 
responsible for the transverse single-spin asymmetries (SSAs) observed in pion
production by the E704 collaboration already some time ago. These data are now 
confirmed, at much higher energy, by the STAR collaboration at 
RHIC~\cite{Rakness}.
A discussion of SSAs in proton-proton collisions (with one proton 
transversely polarized), for both pion production and Drell-Yan processes,
within a perturbative QCD factorization scheme including 
parton transverse motion, was presented by Umberto D'Alesio \cite{dalesio}.
Jacques Soffer \cite{soffer2} discussed SSAs in gauge boson production and  
derived an inequality involving single and double transverse spin asymmetries,
$A_{NN}$ and $A_N$, which might lead to a useful phenomenological bound
for $A_N$.

The origin of single spin asymmetries has recently been much debated using 
various models for the Sivers and Collins functions, relating 
spin and intrinsic motion in, respectively, the distribution of quarks 
inside a transversely polarized proton and the fragmentation of a 
transversely polarized quark into a pion. In particular the Sivers effect, 
which was thought to violate QCD time-reversal invariance, is by now accepted 
as a properly defined phenomenon \cite{col}, with explicit model calculations 
being available \cite{bro}. Some important features of these SSA mechanisms, 
like the factorizability, the universality and the QCD evolution remain a 
challenging open problem, although considerable progress has been made; a 
nice discussion of many of these questions can be found in the contribution 
of Andreas Metz \cite{metz}.  

\section{Perspectives}

The field of high-energy spin physics has constantly grown both in terms of its
theoretical interest, and the number of dedicated experiments and surprising 
results that became available in the last 15 years. As is well-known, spin is 
such a subtle -- truly relativistic and quantum-mechanical -- quantity that 
spin observables provide severe crucial tests of any theory or theoretical 
model. The QCD spin structure of the 
nucleon is a fascinating subject, in that the very simple perturbative QCD 
spin dynamics (helicity conservation) does not match most of the observed spin 
effects. Hence, it is clear that in due time all aspects of QCD spin have to 
be understood.

In the coming years many more data on the spin-structure of the nucleon
will become available. The gluon polarization will be measured by both
COMPASS and RHIC-spin with a fairly good precision, while various data on
exclusive reactions (to study GPDs) will be produced by HERMES and JLab.
The subject of transversity distributions will be addressed by both COMPASS
and HERMES at two different values of $Q^2$, thus making it possible to
verify the predicted weak $Q^2$ dependence by combining both data sets.
Together, all these data will make it possible to quantify the role of
the various carriers of angular momentum in the proton, and provide 
novel tests of QCD in a so-far unexplored domain.

The data will need theoretical interpretation and understanding, aiming at
achieving a consistent definite dynamical picture of nucleon structure,
capable of accounting for any spin effect. Relevant progress has been
made along this line, but many more steps have to be taken: concepts like the 
Generalized Parton Distributions and the Transversity Distribution have to
translate into quantitative experimental observations, giving the necessary
initial non-perturbative input to QCD perturbation theory. QCD evolutions, 
universality and factorization properties have to be discussed and studied,
before it can be claimed that a consistent unified picture has been obtained.
    
\section*{Acknowledgements} We thank all participants of the
working group on spin physics for their contributions 
to the DIS'03 Proceedings. One of the authors (GvdS) is supported by the 
Dutch Foundation for "Fundamenteel Onderzoek der Materie" (FOM).

\end{document}